\documentclass[conference,a4paper]{IEEEtran}
\IEEEoverridecommandlockouts
\usepackage{cite}
\usepackage{amsmath,amssymb,amsfonts}
\usepackage{algorithmic}
\usepackage{graphicx}
\usepackage{textcomp}
\usepackage{array}
\usepackage[caption=false,font=normalsize,labelfont=sf,textfont=sf]{subfig}
\usepackage{epsfig} % for postscript graphics files
\def\BibTeX{{\rm B\kern-.05em{\sc i\kern-.025em b}\kern-.08em
    T\kern-.1667em\lower.7ex\hbox{E}\kern-.125emX}}
\usepackage{geometry}
\begin{document}

\title{Understanding the Temporal Fading in Wireless Industrial Networks: Measurements and Analyses
\thanks{This research is supported by the National Science and Technology Major Project of China under grant 2017ZX03001015, the Scientific Instrument Developing Project of the Chinese Academy of Sciences with Grant No. YJKYYQ20170074, the Natural Science Foundation of Shanghai under grant 18ZR1437500, and the Hundred-Talent Program of Chinese Academy of Sciences under grant Y86BRA1001. The work of T. H. Loh was part-supported by 
The 2017-2020 National Measurement System Programme of the UK government’s Department for Business, Energy and Industrial Strategy (BEIS), under Science Theme Reference EMT17 of that Programme, 
and part-supported by the European Metrology Programme for Innovation and Research (EMPIR) Project 14IND10 (MET5G). The EMPIR is jointly funded by the EMPIR participating countries within EURAMET and the European Union.}
}

\author{\IEEEauthorblockN{1\textsuperscript{st} Qilong Zhang}
\IEEEauthorblockA{\textit{School of Electronics, Electrical}\\ 
\textit{and Communication Engineering} \\
\textit{University of Chinese Academy of Sciences}\\
Beijing, China \\
email:zhangqilong@ucas.ac.cn}

\and
\IEEEauthorblockN{2\textsuperscript{nd} Qiwei Zhang}
\IEEEauthorblockA{\textit{School of Electronics, Electrical}\\ 
\textit{and Communication Engineering} \\
\textit{University of Chinese Academy of Sciences}\\
Beijing, China\\
email:zhangqiwei17@mails.ucas.ac.cn}
% \and
% \IEEEauthorblockN{3\textsuperscript{rd} Xueweu Dai}
% \IEEEauthorblockA{\textit{State Key Lab. of Synthetical}\\
% \textit{ Automation for Process Industries} \\
% \textit{Northeastern University}\\
% Shenyang, China }
\and
\IEEEauthorblockN{3\textsuperscript{rd} Wuxiong Zhang}
\IEEEauthorblockA{\textit{Shanghai Institute of Microsystem}\\ 
\textit{and Information Technology} \\
\textit{Chinese Academy of Sciences}\\
Shanghai, China\\
email:wuxiong.zhang@wico.sh}
\and
\IEEEauthorblockN{4\textsuperscript{th} Fei Shen}
\IEEEauthorblockA{\textit{Shanghai Institute of Microsystem}\\ 
\textit{and Information Technology} \\
\textit{Chinese Academy of Sciences}\\
Shanghai, China\\
email:fei.shen@mail.sim.ac.cn}
\and
\IEEEauthorblockN{5\textsuperscript{th} Tian Hong Loh}
\IEEEauthorblockA{\textit{Time, Quantum \& Elec. Division} \\
\textit{National Physical Laboratory}\\
London, UK \\
email:tian.loh@npl.co.uk}

\and
\IEEEauthorblockN{6\textsuperscript{th} Fei Qin*}
\IEEEauthorblockA{\textit{School of Electronics, Electrical}\\ 
\textit{and Communication Engineering} \\
\textit{University of Chinese Academy of Sciences}\\
Beijing, China\\
email:fqin1982@ucas.ac.cn}
}

\maketitle

\begin{abstract}
The wide deployment of wireless industrial networks still faces the challenge of unreliable service due to severe multipath fading in industrial environments. Such fading effects are not only caused by the massive metal surfaces existing within the industrial environment but also, more significantly, the moving objects including operators and logistical vehicles. As a result, the mature analytical framework of mobile fading channel may not be appropriate for the wireless industrial networks, especially the majority fixed wireless links. In this paper, we propose a qualitative analysis framework to characterize the temporal fading effects of the fixed wireless links in industrial environments, which reveals the essential reason of correlated temporal variation of both the specular and scattered power. Extensive measurements with both the envelop distribution and impulse response from field experiments validate the proposed qualitative framework, which will be applicable to simulate the industrial multipath fading characteristics and to derive accurate link quality metrics to support reliable wireless network service in various industrial applications.
\end{abstract}


\section{Introduction}
Wireless networking has been introduced into industry control systems (termed as wireless industrial network in this paper)~\cite{Gungor2009} due to the advantage of cable-free deployment. The recently proposed Industry 4.0, as well as its variations, heavily rely on the wireless networking to communicate among distributed industry sensors and actuators in a real-time fashion. The most challenging application of wireless industrial networks is the wireless closed-loop control system~\cite{Gao2008a}, which, if viewed from the control side, is a distributed sensing and feedback controlling system “closed” through a wireless network. In these applications, the wireless networks must provide strict constraints on the reliability~\cite{Dai2013} and timing (e.g. delays)~\cite{Huang2015} performance.

The obvious challenge of wireless industrial networks is to deliver guaranteed reliable service in each wireless link when exposed to the multipath fading effect in industrial environments. The deployment locations of wireless industrial networks are usually surrounded by massive metal surfaces with high reflection rate, and therefore cause severe multipath effects. Thus, it is believed that the reliability of wireless networks is highly correlated with the varying multipath fading channel~\cite{Soret2010}. The deployment of wireless nodes in industry location has to first satisfy the application requirements, i.e., usually very close or even directly attached to the metal surface, which would unavoidably affect the radiation pattern and hence the overall link performance of the relevant wireless nodes~\cite{Loh2012}. This deployment habit will not only cause severe multipath due to high reflection rate, and moreover, lead to the fact of fixed wireless link with a main Line of Sight (LOS) path in most scenarios of wireless industrial networks. %This fixed wireless link is always with a main Line of Sight (LOS) path, as no one will intentionally deploy a transmission link cannot see each other\footnote{Although the wireless links with nodes hops away may be blocked by machines or wall, a node in the wireless industrial network is usually be able to see at least one node with LOS path.}.
As the mobile wireless links can be well characterized with the classical mobile fading model in higher fading degree~\cite{Vinogradov2015}, this paper will focus on the majority fixed wireless links in industrial environment.

\begin{figure}
\centering
\includegraphics[width=0.45\textwidth]{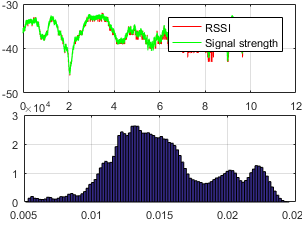}
\caption{The scheme of temporal fading effect in wireless industrial networks}
\label{fig_1}
\end{figure}

The researches of multipath fading in industry scenarios last over decades, most of which are measurement oriented~\cite{Rappaport1989,Hashemi1994,Tanghe2008,Agrawal2014,Vinogradov2015,Cheffena2016a,Eriksson2016,Xu2017}. Although the fading effect in wireless industrial network shares the same principles as the cellular network, the industrial fading effect focused in this paper does show quite different dedicated variation patterns and fails to form a statistical stationary process, as shown in Fig.~\ref{fig_1}. Without any doubts, these fading effects have been already observed many times and termed as the temporal fading effect in~\cite{Hashemi1994,Agrawal2014,Cheffena2016a,Eriksson2016}. Most industry scenarios are usually with the moving objects including nearby operators, logistical vehicles, and gantry cranes etc. The reflection from these moving objects will generate varying paths comparable or even stronger to the LOS path, which will cause observed dedicated fading pattern. However the previous work failed to derive a mature analytical model to discuss the essential reason of temporal fading effect. 
In the other approach, some researchers attempted to formulate a analytical model of multipath fading channel with moving scatters in~\cite{Andersen2009,Zhao2016} following the famous Jakes model~\cite{Aulin1979a}. But most of these work only focus on the Doppler spectrum and avoid the discussion of the envelop fading process. These work argued that even with moving scatters, the caused fading distribution will still follow the Rician distribution. This hypothesis can only be true with large number of moving scatters, which has been demonstrate to be not valid in most field measurements.

%In ideal scenarios of industry, although the delay and attenuation of signals are still varying among multiple paths, the combination of signals over the multipath channel will be static over time. However, most industry scenarios are non-ideal due to the moving objects including nearby operators, logistical vehicles, and gantry cranes etc. The reflection from these moving objects will generate "stronger" paths just less than the LOS path in both amplitude and delay. Therefore, it is straightforward to make the hypothesis that the impulse response (also known as the Power Delay Profile, PDP) should consist of three-layer components. As shown in Fig. 1: the first layer is a static LOS path, the second layer is several varying paths caused by moving objects, and the third layer is several stationary scattered paths with lower amplitude.  Among which, the second layer is caused by only limited number of moving objects with the deterministic pattern, named as temporal fading in previous works~\cite{Hashemi1994}, which is hard to be statistically analyzed. Consequently, it requires a novel framework to analyze, and more essentially, to utilize the characters of multipath effects to provide a reliable network service for the wireless industrial application.

In this paper, we first theoretically analyzed and extended the classical Rician model into temporal fading channel by involving the time varying paths in both the specular and scattered component. To validate the efficiency of proposed framework, a series of experiments have been designed and implemented in industrial environments through Software Defined Radio (SDR) and Vector Network Analyzer (VNA) to observe actual multipath fading effect in both envelop distribution and impulse response. The analysis of these measurements reveal an essential nature of the temporal fading effects in wireless industrial networks, where not only the specular components are time varying due to the nearby moving objects, the scattered components are also time varying in a correlated pattern. The cross-analysis from both the envelop domain and impulse response domain hinted the correctness of proposed qualitative analysis framework for the temporal fading effects. This qualitative framework are expected to support further analytical modeling work, and more essentially, to utilize the characters of temporal effects to provide a reliable network service for the wireless industrial application.

The organization of this paper is as follows: section II provides the qualitative analysis framework, followed by the experiment configuration in section III, as well as the analysis of results discussed in section IV. The conclusion and future works are discussed in section V.

\section{Problem formulation}

The Rician channel model is the most similar scenario with the temporal fading channel in industrial environment, yet has been chosen as the basic model of potential analysis framework. The Rician channel model was proposed to describe the LOS scenario in mobile wireless channel, where the PDP is contributed by a two layer model with a dominated LOS path and several stationary scattering paths~\cite{Aulin1979a,Stuber2012}:
\begin{equation}
\begin{aligned}
\label{eq:5}
h(t)&=C_0e^{j[2\pi f_{D}cos(\theta)t+\varphi_0]}+\sum_{n=1}^NC_ne^{j\varphi_n}\\
%&=\sqrt[]{\frac{K\Omega}{K+1}}h_{sp}(t)+\sqrt[]{\frac{\Omega}{K+1}}h_{sc}(t),
\end{aligned}
\end{equation}
where $C_0e^{j[2\pi f_{D}cos(\theta)t+\varphi_0]}$ is usually termed as the specular component contributed by the LOS path; $\sum_{n=1}^NC_ne^{j\varphi_n}$ is termed as the scattered component contributed by all other scattering paths. %The power of these components have been normalized with $\Omega$ and $K$, where $\Omega$ is the mean power of $h(t)$, $K$ is the Rician factor represent the ratio between specular component and scatter component. 
If assuming the number of scattering paths is large enough, the combination of these scattered component can be generally assumed as a zero-mean complex Gaussian wide-sense stationary random process.
With simple algebra, the envelop of $|h(t)|$ can be derived to follow the Rician distribution $\mathcal{R}(s,\sigma)$~\cite{Aulin1979a}, where $s^2$ equals to the mean power of specular component, and $2\sigma^2$ equals to the mean power of scattered component and will enlarge the error performance.

\begin{figure}
\centering
\includegraphics[width=0.45\textwidth]{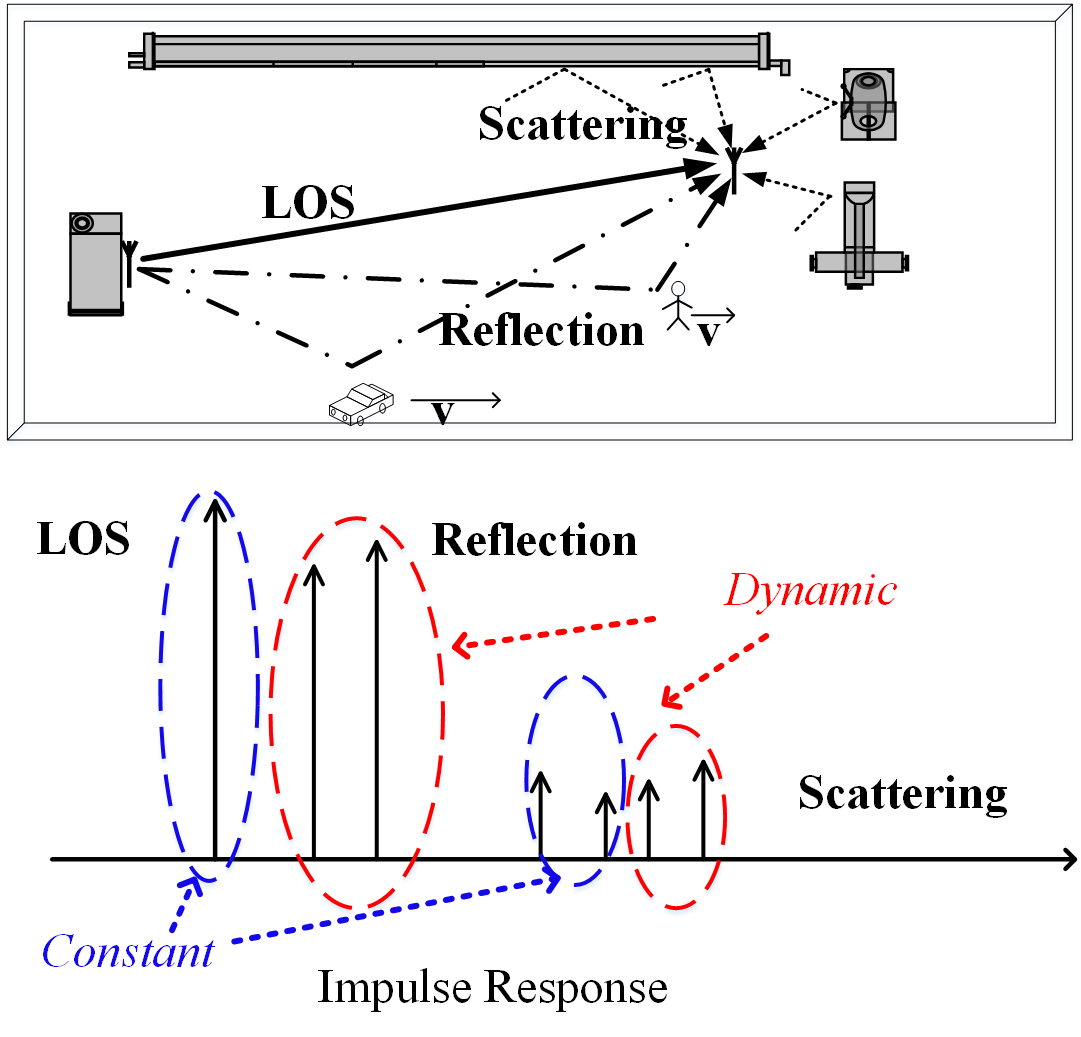}
\caption{ The scheme of impulse response of temporal fading effect in wireless industrial networks.}
\label{fig_ir}
\end{figure}

In the scenarios of fixed wireless link in wireless industrial networks, the phase term in equation (\ref{eq:5}) should be rewritten by avoiding the Doppler effect and involving the propagation delay: %$\Phi_n=\varphi_n-2\pi f_c\tau_n$.
\begin{equation}
\label{eq:6}
\Phi_n=\varphi_n-2\pi f_c\tau_n.
\end{equation}
This reveals the fact that the phase term of each path will no longer vary with time but with the propagation delay. This effect has can be validated by the intuitive observation: a fixed wireless link, even in the severe multipath scenario, will only suffer from a static flat fading. However, the fixed wireless link in the industrial networks will suffer from another, even worse, challenge caused by the nearby moving objects. As shown in Fig.~\ref{fig_ir}, the nearby moving objects, like human operators, logistical vehicles, and gantry cranes etc., will cause variations to both the $C_n$ and $\tau_n$ of signal paths reflected by them. Under the consideration that the scattered paths are essentially the high order results after many times of reflection, scattering, and diffractions, the corresponding scattered paths will also shown such variations but in a much lower ratio. Hence, it is straightforward to rewrite equation (\ref{eq:5}) as:
\begin{equation} 
\begin{aligned}
\label{eq:7}
h(t)&=C_0e^{j\varphi_0}+\sum_{n=1}^{N_r}{C_n(t)e^{j[\varphi_n-2\pi f_c\tau_n(t)]}}\\
&+\sum_{n=N_r+1}^{N_s}C_ne^{j\varphi_n} + \sum_{n=N_s+1}^NC_n(t)e^{j\varphi_n},
\end{aligned}
\end{equation}
where $C_0e^{j\varphi_0}$ is the static LOS path. The second item $\sum_{n=1}^{N_r}{C_n(t)e^{j[\varphi_n-2\pi f_c\tau_n(t)]}}$ are the reflected paths with comparable amplitudes to the LOS path. These paths are configured with the time varying $C_n(t)$ and $\tau_n(t)$ to represent the effect from the moving nearby objects with dedicated pattern. Beware that the number of these paths is usually limited, yet $C_n(t)$ and $\tau_n(t)$ failed to form a statical random process and should be modeled as a dedicated dynamic function. The 3rd item $\sum_{n=N_r+1}^{N_s}C_ne^{j\varphi_n}$ refers the scattered paths caused by LOS path, while the 4th item $\sum_{n=N_s+1}^NC_n(t)e^{j\varphi_n}$ represents the scattered paths which can be rooted to the moving of nearby objects, i.e. also with dedicated varying amplitude $C_n(t)$. As the scattered paths are the signals reflected several times, the phase $\varphi_n$ in both the 3rd and 4th item usually exhibits a stationary uniform distribution. 
With simple algebra and normalization, the first two items and later two items can be combined with their general form respectively. Then, equation (\ref{eq:7}) could be rewritten as:
\begin{equation}
\begin{aligned}
\label{eq:8}
h(t)=&\sum_{n=0}^{N'}{C_n(t)e^{j[\varphi_n-2\pi f_c\tau_n(t)]}}+\sum_{n=N'}^NC_n(t)e^{j\varphi_n}.
\end{aligned}
\end{equation}

Equation (\ref{eq:8}) shows a similar pattern of the standard Rician model but with dynamic power of both specular and scattered components. If further considering that the velocities of moving objects in industry scenarios are usually low, which can be demonstrated by the slow variation of signals in Fig.~\ref{fig_1}, i.e. usually in the level of several tens of seconds. Then, within a frame/packet duration, i.e. usually in the level of few milliseconds, $C_n(t)$ and $\tau_n(t)$ can be roughly assumed to be static. Thus, the Power Density Function (PDF) of received amplitude at any observing time can be expressed as a static Rician distribution $\mathcal{R}(s,\sigma)$ with $s$ contributed by $\sum_{n=0}^{N'}{C_n(t)e^{j[\varphi_n-2\pi f_c\tau_n(t)]}}$ and $\sigma$ contributed by $\sum_{n=N'}^NC_n(t)e^{j\varphi_n}$.If viewed globally, the temproal fadding channel can be described by a dynamic Rician distribution with contiuesly time varying parameters $\mathcal{R}(s(t),\sigma(t))$. Clearly, this will lead to three essential facts of the temporal fading channel: 
\begin{enumerate}
\item both mean power and variation power will be time varying due to the movement of nearby objects; 
\item this temporal varying failed to form a stationary distribution; 
\item the mean power and variation power will be varying in a correlated pattern; 
\end{enumerate}

\section{Measurement setup}
In the experiment, a ZNB-8 Vector Network Analyzer from R\&S has been employed to obtain the real-time impulse response with sample rate of 200$ms$, while two NI USRP-2922 SDR transceivers have been employed to obtain the envelope distribution of the received signal. We try to observe the variation of envelope of received signals through the I/Q stream from SDR platform~\cite{Loh2012}. The transmitted signals are configured with Quadrature Phase Shift Keying (QPSK) modulation with randomized bit stream, instead of the wide utilized Continue Wave (CW) signals in many reported work. Under this configuration of the experiment, the findings will not only be analyzed to help understanding the industrial fading channel, but also be able to be straightforwardly extend to the next research stage, i.e. generate an accurate link quality metric in industrial fading channel through baseband I/Q stream. The raw ADC reading of I/Q stream has been normalized with the help of a spectrum analyzer to make the conversion of $I^2+Q^2$ just in the unit of $mW$ and its variation $dBm$. In the experiment, each estimation is based on the per packet manner which means the "sampling point" in x-axis refers to 5$ms$ in time.

\begin{figure}
\centering
\includegraphics[width=0.45\textwidth]{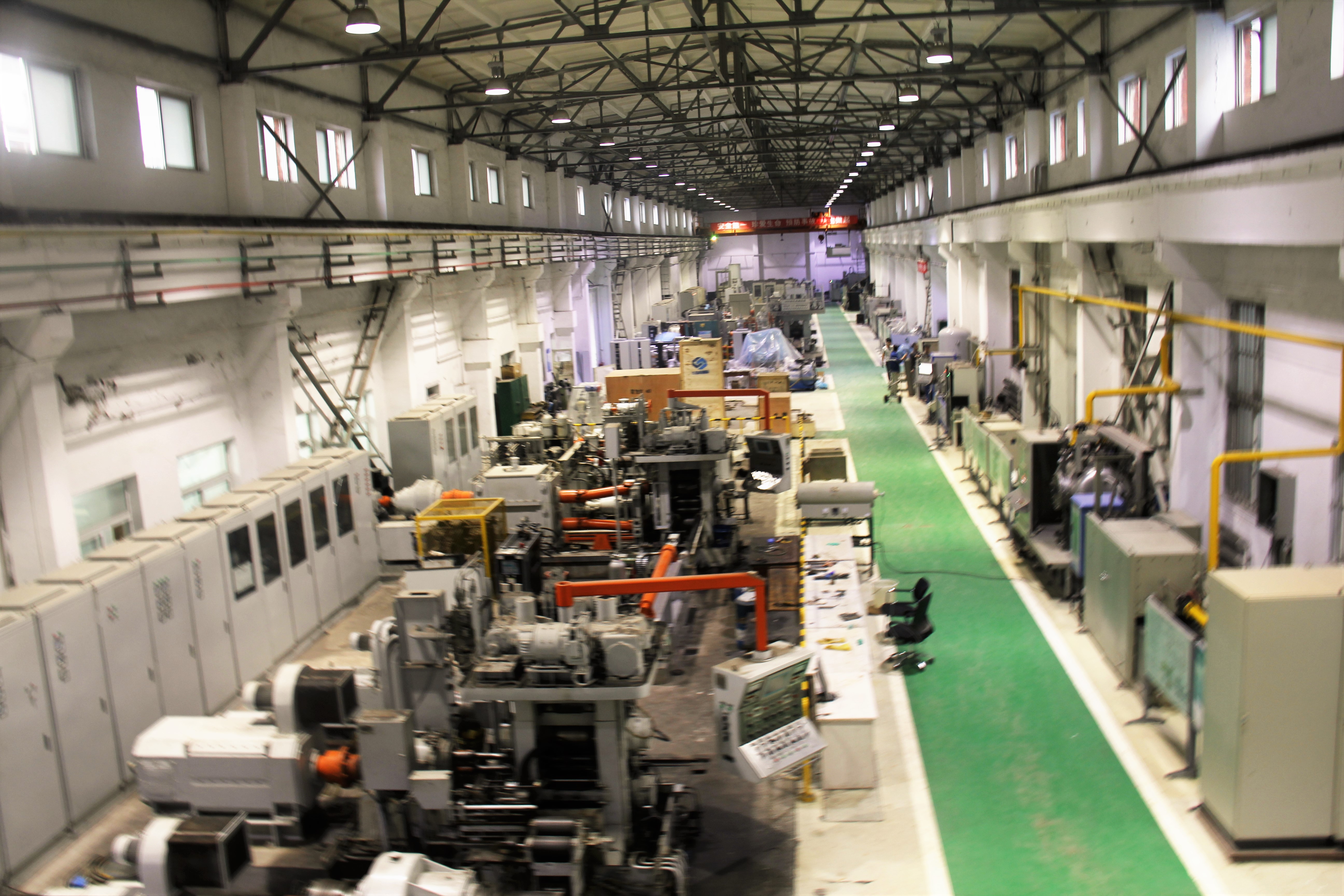}
\caption{Details of experiments environment}
\label{fig_2}
\end{figure}

All the experiments are equipped with 2.5 dBi omni-directional antennas, type CS900/1800-3 from Jingcheng Electronics. The antenna was connected to measurement device through a 5 meters RF cable, type Drake RG223, which provides good resistance to the bend of cable. The antenna was attached to the surface of machines to emulate the general operation of the wireless industrial network. The carrier frequency of experiment was set to be the popular 915 MHz in industry applications by default, and in some case 2400 MHz for comparison~\cite{Tanghe2008}. Although CS900/1800-3 was designed to work in 900 MHz and 1800 MHz, the initial tests of the reflection coefficient (i.e. antenna operational bandwidth) using Vector Network Analyzer demonstrated good matching around 2400 MHz as well. Thus, all experiments were measured without the change of antennas.

As shown in Fig.\ref{fig_2}, most field measurements were obtained in a rolling mill around $300\times20\times10$ meters large, with a big gantry crane near the roof. 
% Experiments were carried out near a rolling line, 
% %as shown in Fig.~\ref{fig_2}, 
% with the transmitting antenna at the start location and receiving antenna at the end of rolling line, both attached to the motionless surface of machines. The carrier frequency of USRP was set to be the popular 915 MHz in industry applications by default and in some case 2400 MHz for comparison. 
Some experiments, e.g. the impulse response test and no moving objects test, were deployed in an emulation teaching lab also with massive metal surfaces, % shown in Fig. 2 (c), 
where the ideal scenarios can be emulated. The third series of experiments were carried with the same devices pair connected through a radio channel emulator of keysight Propsim F8, which can be configured with manually input PDP, and are utilized for cross-validation.
%The rolling mill is under operation in the deployed experiment, thus the experiment time was carefully selected to avoid the Electro-Magnetic (EM) interference caused by the steel cutter etc., as the solution for both EM interference and other RF interference (WiFi etc.) is out of the scope for this paper, which can be referenced to the existed work of ~\cite{Qin2013}.

\section{Analysis of Measurement Results}

\subsection{Envelop distribution}

\begin{figure}
%generates subfigures with title "a,b...",instead of "fig.1"
\centering
\subfloat[]{\includegraphics[width=0.45\textwidth]{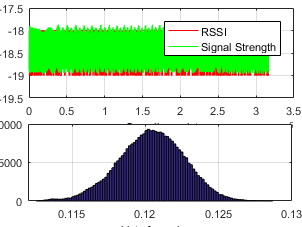}
\label{fig_3_a}}
\hfil
\subfloat[]{\includegraphics[width=0.45\textwidth]{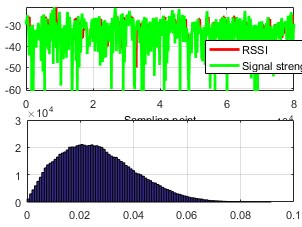}
\label{fig_3_b}}
\hfil
% \subfloat[]{\includegraphics[width=0.4\textwidth]{fig7}
% \label{fig_6_b}}
\caption{Received Envelop with classical channels: (a)AWGN channel; (b)Rician channel.}
\label{fig_3}
\end{figure}

In this subsection, the envelop distributions of the received signal will be presented first. 
%These experiments were implemented to validate the efficiency of employed devices. 
The results in Fig. \ref{fig_3} (a) were obtained through direct cable connection between transmission pair, which leads to a classical AWGN channel with only internal thermal noise. The received signal strengths are provided in two forms: the red curve is the Received Signal Strength Indicator (RSSI) estimation, i.e. a 16$\mu s$ segment at the beginning of each 5$ms$ frame will be averaged and rounded to generate the RSSI metric (aided by the known RX amplifier configuration, this algorithm equals to the analog estimation in the front-end). The green curve is the 'true' value of received signal strength in dBm for each symbol. The histogram amplitude of received signal in amplitude has been further provided in the lower subplot of the figure, which shows very good fits of Gaussian distribution. Similar experiment results from the radio channel emulator configured with a classical Rician channel has been provided in Fig. \ref{fig_3} (b). Both the received signal strength and histogram show good fits to the Rician distribution. The results shown in Fig. \ref{fig_3} not only demonstrate the feasibility of employing baseband I/Q stream to estimate channel effect, but also reveals the fact that commonly utilized RSSI estimation shows good performance in AWGN channel, but fail to satisfy the utilization in varying fading channel. For the samples in Fig. \ref{fig_3} (b), an average estimation error of 7.5 dB, as well as a maximum error of 31.5 dB, have been noticed.

% \begin{figure}
% \centering
% \includegraphics[width=0.4\textwidth]{fig3b}
% \caption{Results with classic channels.}
% \label{fig_3}
% \end{figure}

% %[H]without which images always appear on top of the page
% \begin{figure}
% \centering
% \includegraphics[width=0.4\textwidth]{fig4a}
% \caption{Results with Rician channel.}
% \label{fig_4}
% \end{figure}

\begin{figure}
\centering
\includegraphics[width=0.45\textwidth]{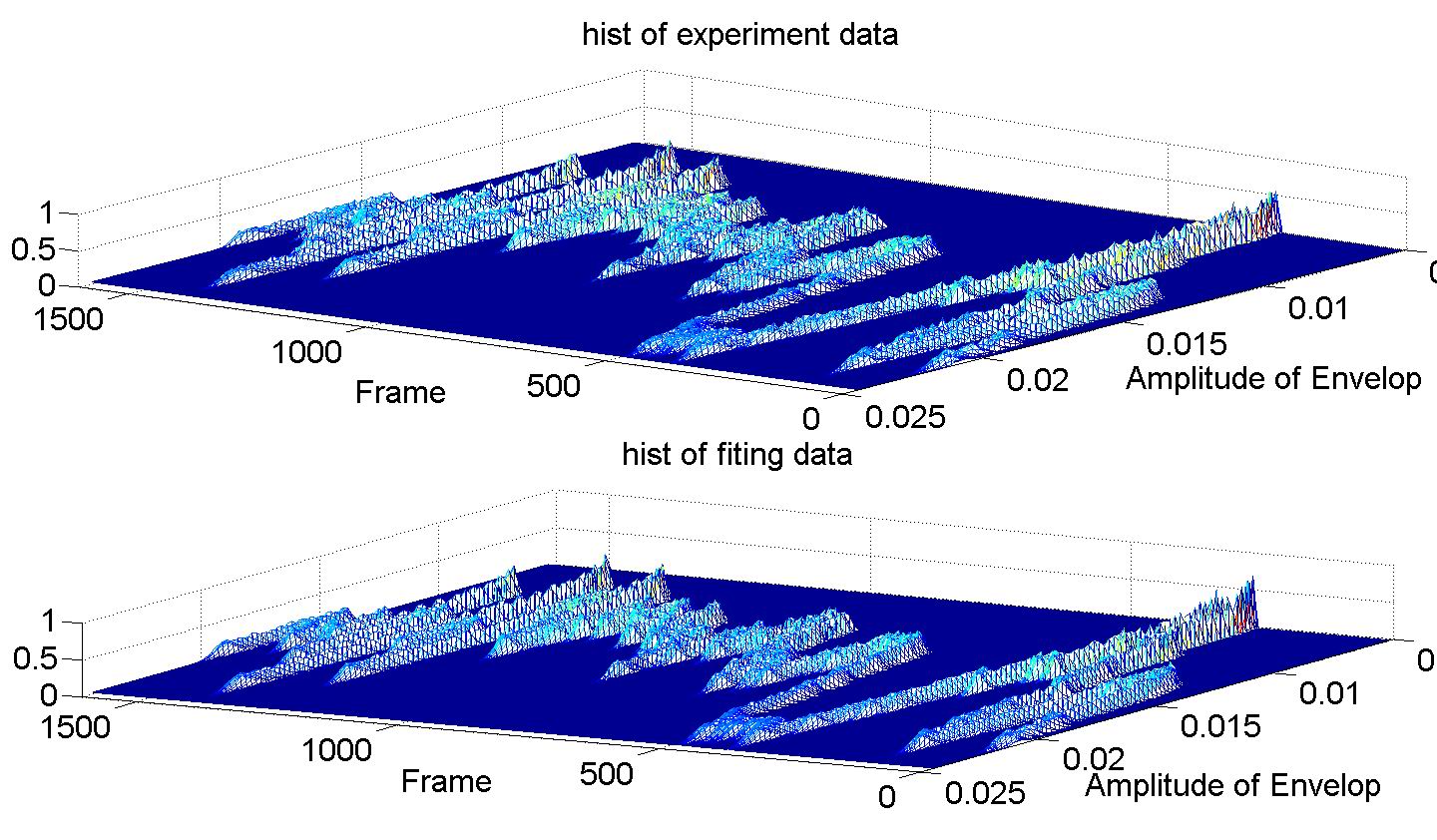}
\caption{Frame sliced histogram of signal envelop in temporal fading channel}
\label{fig_6}
\end{figure}

%According to the discussion in section II, the variation of specular power will be decided by both the LOS path and reflected paths and show the dedicated temporal pattern due the moving of limited reflectors. 

In most filed experiments, a logistical vehicle was employed to moving around the transmission pair. The experiment has been repeating several times with similar finding, yet only a classical example has been provided.
The intuitive results with nearby moving vehicle have been first shown in Fig.~\ref{fig_1}, where the received signal strength varies from -46dBm to -33dBm. The histogram of overall received signals has also been provided but fails to fit into any known distributions. If compared with the classical Rician distribution of mobile channel shown in Fig.~\ref{fig_3} (b), the variation speed of the temporal fading channel is much slower. By assuming the channel variation are caused by the nearby moving object, and considering the moving objects in industry scenario are usually with very low speed, a hypothesis can be proposed that in the small time scale, i.e. a frame length of 5ms, the location of moving object can be assumed to be fixed as well as the envelop distribution. With this hypothesis, the received envelops have been reshaped into frames. The histograms of signal strengths of each frame as well as the Rician distribution fitting results through Maximum Likelihood estimation have been illustrated in Fig.~\ref{fig_6}. The residue of fitting error calculated in PDF with:
\begin{equation}
\begin{aligned}
\label{eq:11}
\frac{1}{F}\sum_{f=1}^F\sum_{k=1}^N\frac{{(r_k-\hat{r_k})}^2}{r_k}
\end{aligned}
\end{equation}
where $f$ is the index of frames. 

The results have been provided in Table 1, where all four experiments with different configurations show very tiny residue error (e.g. the residue error of estimation for Fig.~\ref{fig_6} is 0.4\% termed as '915\_long' in Table 1). The results clearly demonstrate that the histogram of signal strengths inside a frame can be fitted accurately by a dynamic Rician distribution with continues changing parameters $s$ and $\sigma$. 

\begin{table}[h!]
\centering
\caption{The Residue of Fitting Error}
\begin{tabular}{ | m{1.3cm} | m{1.3cm} | m{1.3cm} | m{1.3cm} | m{1.3cm} | } 
\hline
Scenario & 915\_Long & 915\_short & 2.4\_Long & 2.4\_short \\ 
\hline
Estimation error & 0.0040 & 0.0044 & 0.0044 & 0.0031 \\ 
\hline
\end{tabular}
\end{table}

\begin{figure}
\centering
\includegraphics[width=0.45\textwidth]{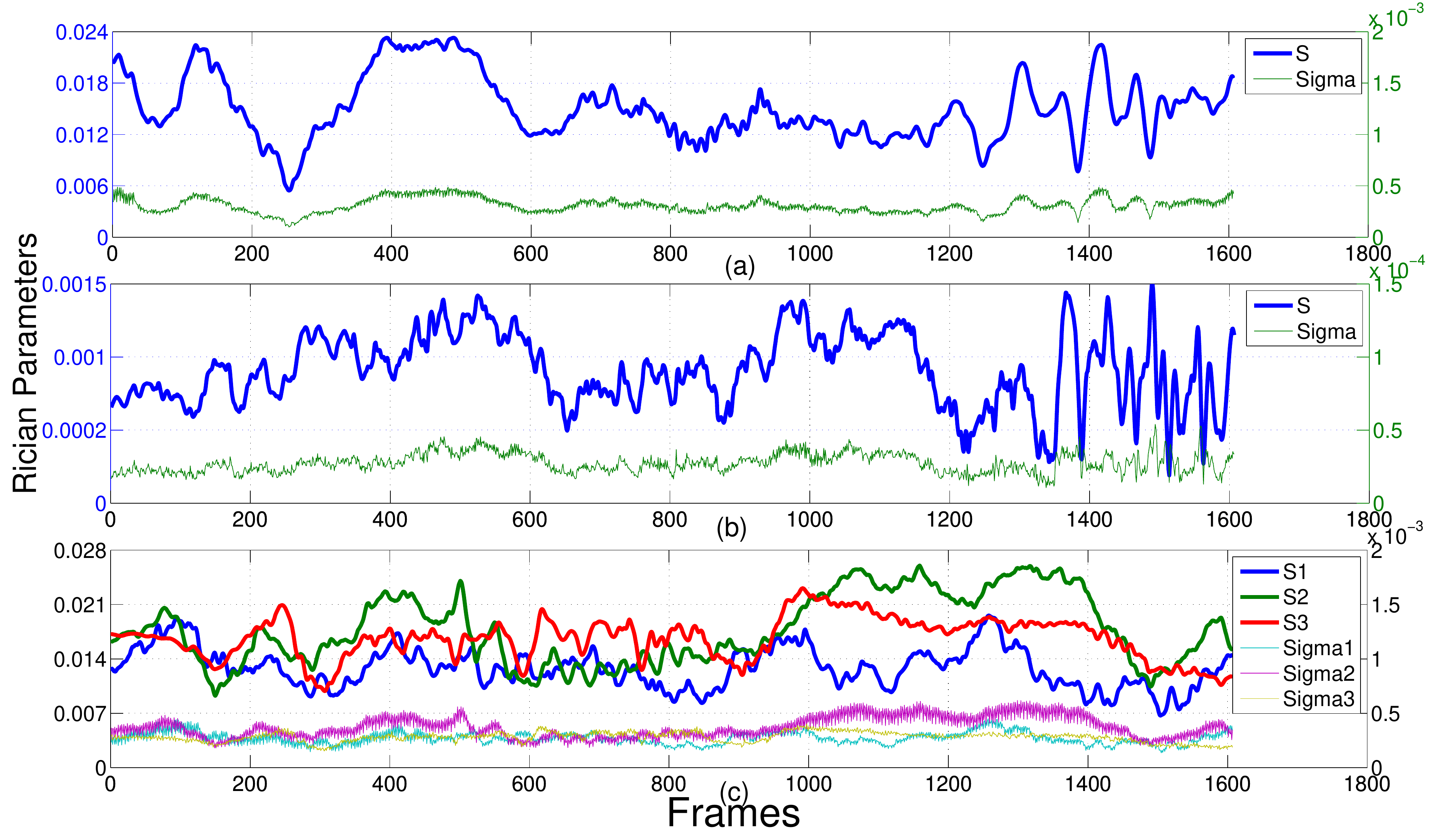}
\caption{The estimated parameters of envelop distribution with sliced frames: (a)results with moving vehicle in 915MHz; (b)results with moving vehicle in 2.4GHz; (c)results with different number of moving humans in 915MHz}
\label{fig_7}
\end{figure}

The estimated parameters of Rician distributions from the same experiment of Fig.~\ref{fig_1} and Fig.\ref{fig_6} along the time stepping have been drawn in Fig.~\ref{fig_7} (a). An important fact has been revealed by Fig.~\ref{fig_7} (a):
easy to notice that $s$ changes around 0.020 with the highest value of 0.024 and lowest value of 0.008, while the same time although $\sigma$ changes around a much lower level around $4 \times 10^{-4}$, it does change in a correlated pattern with $s$. Not surprisingly, the observed phenomenons are hardly to be explained by existing analytical framework even the one proposed for the moving scatters, which usually assumes the stationary distribution of scattered components. In the other side, the proposed qualitative framework perfectly fits these measurement results with the involvement of time varying paths and propagation distance based phase.  

To further validate such finding, additional experiments have been designed. The experiment results obtained with 2.4 GHz have been provided in Fig.~\ref{fig_7} (b), which show the similar effect but higher jitter effect due to the higher carrier frequency. This validates the proposed framework is applicable for various frequency. The measurements with the moving human body, i.e. the worker or operators, have been presented in Fig.~\ref{fig_7} (c). In these experiments, one, two, and three operators moving irregularly near the fixed wireless link. All the three $s$ change around 0.014 with highest to 0.026 and lowest to 0.008 all with the correlated varied $\sigma$, which are almost the same with the results of moving vehicle. The similar results of these three scenarios validate the proposed framework works for both the metal surface with high reflection rate and objects like a human body with low and irregular reflection rate. 

As reported in \cite{Hashemi1994}, although the effect of single or few moving objects is non-stationary, the combination of these effects may contribute a stationary process when the number of objects is large enough. However, the experiment shown in Fig.~\ref{fig_7} (c) fails to demonstrate this hypothesis, i.e. all three scenarios show similar non-stationary effect. This may due to the fact that three people may not enough, especially in the large rolling mill space. We will try to examine this issue in our future work to see whether a stationary process caused by a large number of nearby objects could be located.

\subsection{Impulse Response}

\begin{figure}
%generates subfigures with title "a,b...",instead of "fig.1"
\centering
\subfloat[]{\includegraphics[width=0.45\textwidth]{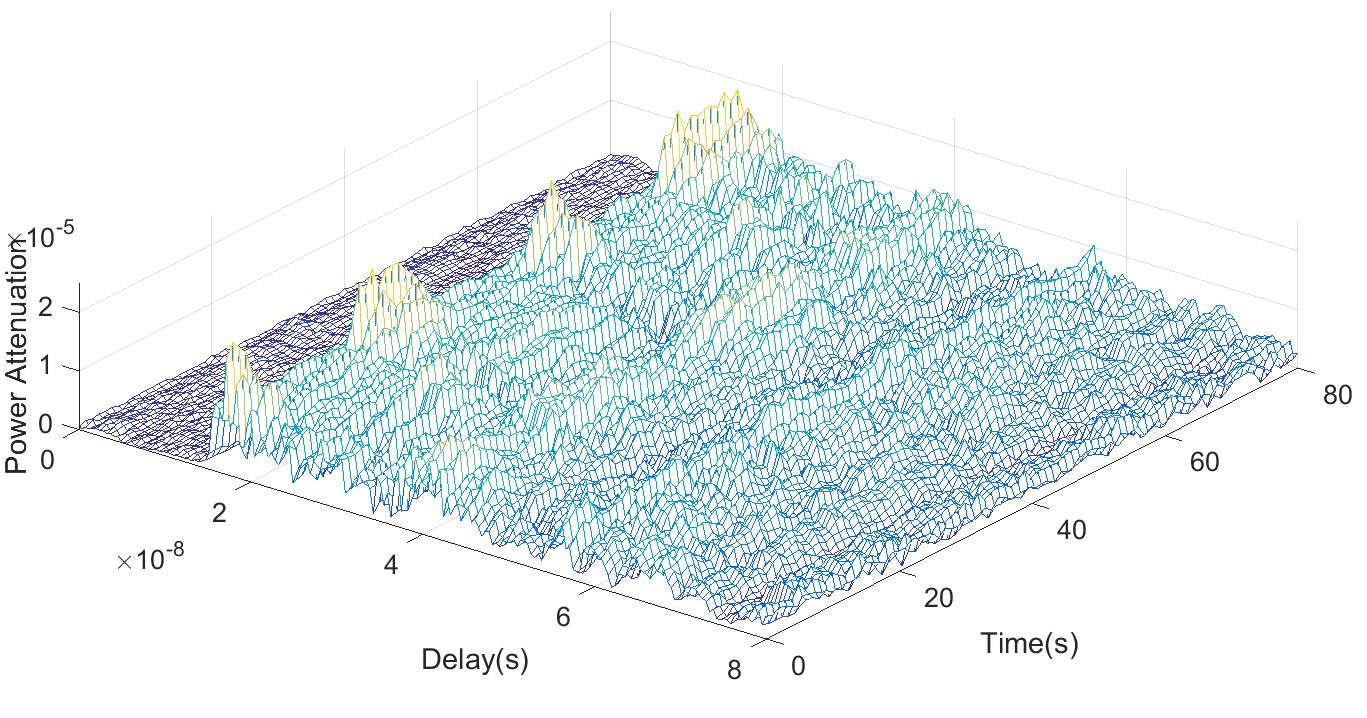}
\label{fig_6_a}}
\hfil
\subfloat[]{\includegraphics[width=0.45\textwidth]{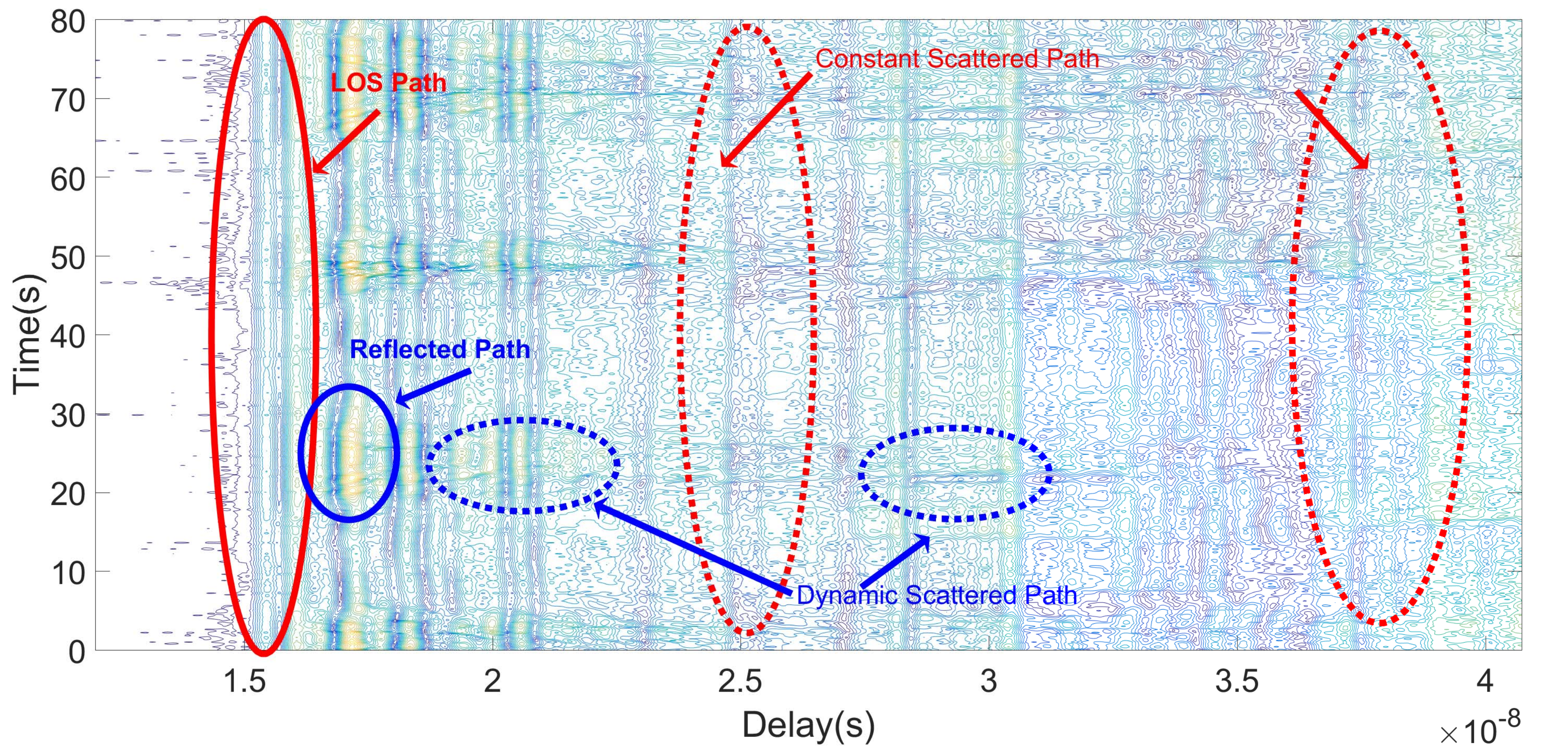}
\label{fig_6_b}}
\hfil
\subfloat[]{\includegraphics[width=0.45\textwidth]{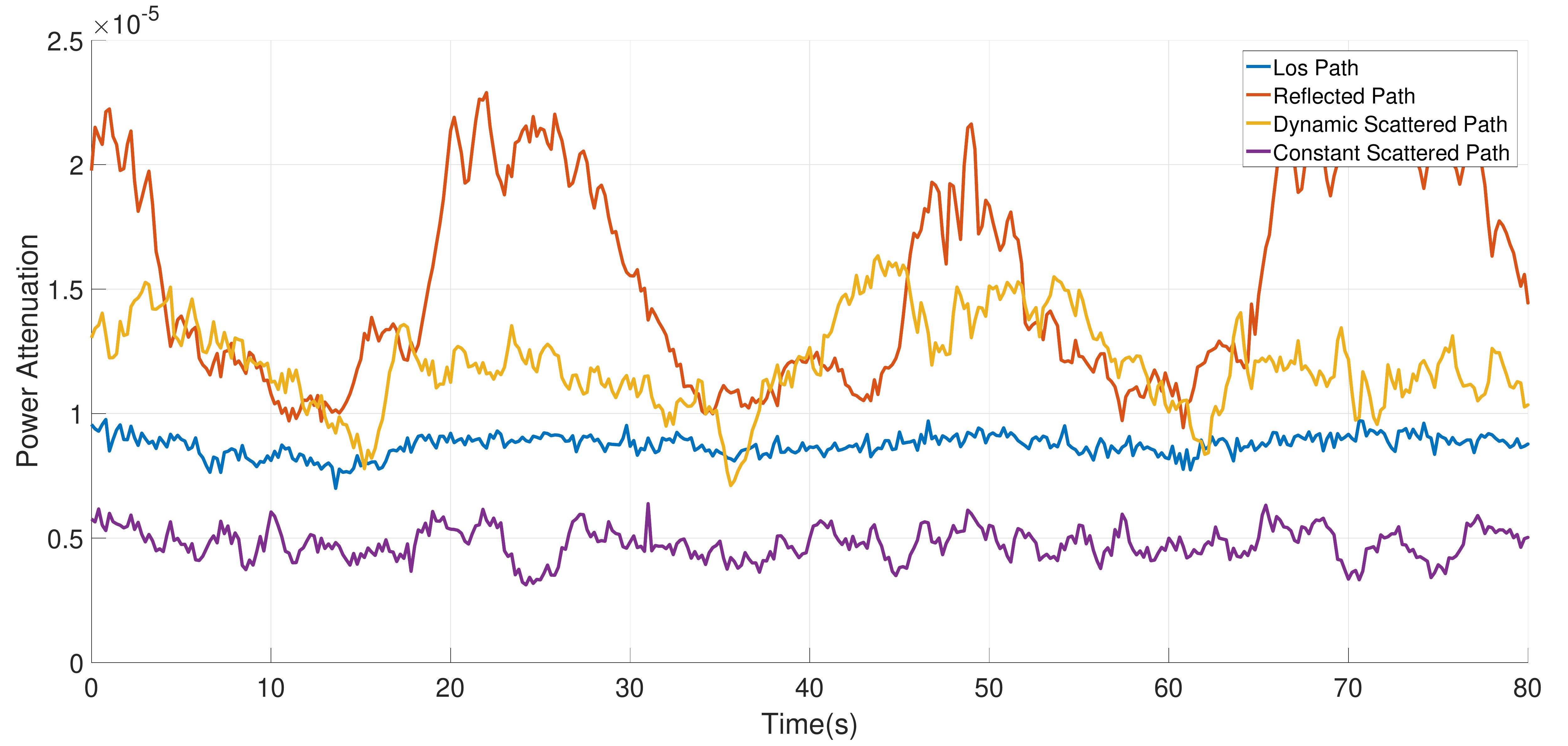}
\label{fig_6_c}}
\hfil
% \subfloat[]{\includegraphics[width=0.4\textwidth]{fig7}
% \label{fig_6_b}}
\caption{Impulse Response results: (a)Impulse Response in time series; (b)Amplified local detail in contour;  (c)The typical variation patterns of different paths.}
\label{fig_PDP}
\end{figure}

% \begin{figure}
% \centering
% \includegraphics[width=0.45\textwidth]{DopplerSpectra}
% \caption{Doppler Spectra results}
% \label{fig_DopplerSpec}
% \end{figure}

In this subsection, the Impulse responses captured with VNA in similar scenarios of Fig.~\ref{fig_1} have been presented to help the cross-analysis of the temporal fading effects. In Fig.~\ref{fig_PDP}.(a), the intuitive impulse response overview has been provided with X axis of impulse delay, i.e. $\tau$, and Y axis of measurement time in the rate of 200$ms$. In the experiment, a logistical vehicle was moving around the fixed wireless link in a regular routing, e.g. from location A to B and vice versa. As a result, the most significant peak has shown the regular variation pattern as well as some less significant peaks have shown the correlated variation pattern. To view this effect quantitatively, the significant area of the impulse responses have been amplified and shown in the contour form of Fig.~\ref{fig_PDP}.(b), as well as the selected typical paths shown in Fig.~\ref{fig_PDP}.(c). 

The LOS path has been first noticed around $\tau=15ns$ with the shortest delay (i.e., $\tau$) and the near constant level (i.e., $C$) according to the theory. The second noticed path is the most significant reflected path around $\tau=17ns$ with variations in both amplitude $C(t)$ and propagation delay $\tau(t)$. The higher level than the LOS path is reasonable, since it could be the combination of several reflected paths with the same path length~\cite{Stuber2012}. Typical scattered paths have been presented as follows. It is then very straightforward to assume that the scattered path are the higher order reflected paths, then the variations in $\tau$ will trend to be negligible. Some scattered paths clearly showing correlated variations in $C(t)$ are termed as dynamic scattered paths, while the scattered paths rooted to the constant LOS path are termed as constant scattered paths. Such measurements are perfectly compatible with the observed dynamic envelop distribution shown in section IV.A, i.e. the LOS path and dynamic significant reflector paths will contribute to the time varying mean power of envelop distribution, while all the other paths will contribute to the variation of envelop distribution. As some of the scattered paths shows correlated pattern with the significant reflector paths, the expected correlated varying pattern in both the mean power and variation power of envelop distribution can also be expected. The combination of both the envelop distribution and impulse finally cross-validate the proposed qualitative analysis framework for the temporal fading channel in industrial environment.

\section{CONCLUSIONS}

The temporal fading effects in the industrial environment have been observed and reported over decades, the unique pattern of which is significantly different with the classical mobile fading channel. To understand the essential natures and reveal the scientific reasons of the temporal fading effect, this paper designed a series of experiments in real industrial environments. The cross-validation of both the time varying envelop distribution and the impulse response has provided, which validates the proposed qualitative analysis framework of the temporal fading channel. In the future work, such a qualitative analysis framework can be further extended to a mature analytical model of temporal fading channel in industrial environment, which are expected to increase the performance of wireless industrial networks.

%This paper proposes a novel framework to analyze the fading effect for the majority fixed wireless links in industrial scenarios. In particular, a dynamic distribution with continuously changing parameters can be employed to characterize the temporal industrial multipath fading, which rooted to a three layers impulse response model. A series of experiments from real industry sites have been designed to validate this hypothesis. To our best knowledge, this is the first time to propose the correlated variation pattern of scattered power with both analyses and measurements, which was usually assumed as stationary in classical wireless channel theory. In the future work, this analytical framework can be further extended to provide a trustable simulation model of industry fading channel as well as an accurate link quality metric to aid reliable transmission in the wireless industrial network, which in turn, is believed to be able to significantly promote the market of wireless networks in industry applications.

% \section*{Acknowledgment}

% The preferred spelling of the word ``acknowledgment'' in America is without 
% an ``e'' after the ``g''. Avoid the stilted expression ``one of us (R. B. 
% G.) thanks $\ldots$''. Instead, try ``R. B. G. thanks$\ldots$''. Put sponsor 
% acknowledgments in the unnumbered footnote on the first page.

\bibliographystyle{IEEEtran}
\bibliography{library.bib}

\begin{thebibliography}{10}
\providecommand{\url}[1]{#1}
\csname url@samestyle\endcsname
\providecommand{\newblock}{\relax}
\providecommand{\bibinfo}[2]{#2}
\providecommand{\BIBentrySTDinterwordspacing}{\spaceskip=0pt\relax}
\providecommand{\BIBentryALTinterwordstretchfactor}{4}
\providecommand{\BIBentryALTinterwordspacing}{\spaceskip=\fontdimen2\font plus
\BIBentryALTinterwordstretchfactor\fontdimen3\font minus
  \fontdimen4\font\relax}
\providecommand{\BIBforeignlanguage}[2]{{%
\expandafter\ifx\csname l@#1\endcsname\relax
\typeout{** WARNING: IEEEtran.bst: No hyphenation pattern has been}%
\typeout{** loaded for the language `#1'. Using the pattern for}%
\typeout{** the default language instead.}%
\else
\language=\csname l@#1\endcsname
\fi
#2}}
\providecommand{\BIBdecl}{\relax}
\BIBdecl

\bibitem{Gungor2009}
V.~Gungor and G.~Hancke, ``{Industrial Wireless Sensor Networks: Challenges,
  Design Principles, and Technical Approaches},'' \emph{IEEE Trans. Industrial
  Electronics}, vol.~56, no.~10, pp. 4258--4265, 2009.

\bibitem{Gao2008a}
H.~Gao, T.~Chen, and J.~Lam, ``{A new delay system approach to network-based
  control},'' \emph{Automatica}, vol.~44, no.~1, pp. 39--52, 2008.

\bibitem{Dai2013}
X.~Dai, Z.~Gao, and G.~Z. {Dai X.}, ``{From Model, Signal to Knowledge: A
  Data-Driven Perspective of Fault Detection and Diagnosis},'' \emph{IEEE
  Transaction on Industrial Informatics}, vol.~9, no.~4, pp. 2226--2238, 2013.

\bibitem{Huang2015}
Y.~Huang, T.~Li, X.~Dai, H.~Wang, and Y.~Yang, ``{TS2: A realistic IEEE1588
  time-synchronization simulator for mobile wireless sensor networks},''
  \emph{Simulation}, vol.~91, no.~2, pp. 164--180, 2015.

\bibitem{Soret2010}
B.~Soret, M.~C. Aguayo-Torres, and J.~T. Entrambasaguas, ``{Capacity with
  Explicit Delay Guarantees for Generic Sources over Correlated Rayleigh
  Channel},'' \emph{Wireless Communications IEEE Transactions on}, vol.~9,
  no.~6, pp. 1901--1911, 2010.

\bibitem{Loh2012}
T.~H. Loh and C.~M., ``{Radiation pattern characterisation of embedded radios
  emitting modulated signals},'' in \emph{19th International Conference on
  Microwave Radar and Wireless Communications (MIKON)}, 2012, pp. 80--84.

\bibitem{Vinogradov2015}
E.~Vinogradov, W.~Joseph, and C.~Oestges, ``{Measurement-Based Modeling of
  Time-Variant Fading Statistics in Indoor Peer-to-Peer Scenarios},''
  \emph{IEEE Transactions on Antennas and Propagation}, vol.~63, no.~5, pp.
  2252--2263, 2015.

\bibitem{Rappaport1989}
T.~S. Rappaport and C.~D. Mcgillem, ``{UHF Fading in Factories},'' \emph{IEEE
  Journal on Selected Areas in Communications}, vol.~7, no.~1, pp. 40--48,
  1989.

\bibitem{Hashemi1994}
H.~Hashemi, M.~McGuire, T.~Vlasschaert, and D.~Tholl, ``{Measurements and
  modeling of temporal variations of the indoor radio propagation channel},''
  \emph{IEEE Transactions on Vehicular Technology}, vol.~43, no.~3, pp.
  733--737, 1994.

\bibitem{Tanghe2008}
E.~Tanghe, W.~Joseph, L.~Verloock, L.~Martens, H.~Capoen, K.~{Van Herwegen},
  and W.~Vantomme, ``{The industrial indoor channel: Large-scale and temporal
  fading at 900, 2400, and 5200 MHz},'' \emph{IEEE Transactions on Wireless
  Communications}, vol.~7, no.~7, pp. 2740--2751, 2008.

\bibitem{Agrawal2014}
P.~Agrawal, A.~Ahlen, T.~Olofsson, and M.~Gidlund, ``{Long term channel
  characterization for energy efficient transmission in industrial
  environments},'' \emph{IEEE Transactions on Communications}, vol.~62, no.~8,
  pp. 3004--3014, 2014.

\bibitem{Cheffena2016a}
M.~Cheffena, ``{Propagation Channel Characteristics of Industrial Wireless
  Sensor Networks [Wireless Corner]},'' \emph{IEEE Antennas and Propagation
  Magazine}, vol.~58, no.~1, pp. 66--73, feb 2016.

\bibitem{Eriksson2016}
M.~Eriksson and T.~Olofsson, ``{On Long-Term Statistical Dependences in Channel
  Gains for Fixed Wireless Links in Factories},'' \emph{IEEE Transactions on
  Communications}, vol.~64, no.~7, pp. 3078--3091, 2016.

\bibitem{Xu2017}
J.~Xu, X.~Yan, Y.~Zhu, J.~Wang, Y.~Yang, X.~Ge, G.~Mao, and O.~Tirkkonen,
  ``{Statistical Analysis of Path Losses for Sectorized Wireless Networks},''
  \emph{IEEE Transactions on Communications}, vol.~65, no.~4, pp. 1828--1838,
  2017.

\bibitem{Andersen2009}
J.~B. Andersen, J.~O. Nielsen, G.~F. Pedersen, G.~Bauch, and G.~Dietl,
  ``{Doppler spectrum from moving scatterers in a random environment},''
  \emph{IEEE Transactions on Wireless Communications}, vol.~8, no.~6, pp.
  3270--3277, 2009.

\bibitem{Zhao2016}
X.~Zhao, Q.~Han, X.~Liang, B.~Li, J.~Dou, and W.~Hong, ``{Doppler Spectra for
  F2F Radio Channels with Moving Scatterers},'' \emph{IEEE Transactions on
  Antennas and Propagation}, vol.~64, no.~9, pp. 4107--4112, 2016.

\bibitem{Aulin1979a}
T.~Aulin, ``{A Modified Model for the Fading Signal at a Mobile Radio
  Channel},'' \emph{IEEE Transactions on Vehicular Technology}, vol.~28, no.~3,
  pp. 182--203, 1979.

\bibitem{Stuber2012}
G.~Stuber, \emph{{Principles of Mobile Communication}}, 3rd~ed.\hskip 1em plus
  0.5em minus 0.4em\relax Springer, 2012.

\end{thebibliography}

\end{document}